\RequirePackage{fix-cm}
\documentclass[natbib,smallextended]{svjour3}       
\smartqed  
\usepackage{graphicx}
\journalname{my journal}
\begin{document}

\title{Dust evolution in protoplanetary discs and the formation of planetesimals}
\subtitle{What have we learned from laboratory experiments?}

\titlerunning{Formation of planetesimals}        

\author{J\"urgen Blum}

\institute{Institut f\"ur Geophysik und extraterrestrische Physik\\  Technische Universit\"at Braunschweig \at
              Mendelssohnstr. 3\\ 38106 Braunschweig \\Germany \\
              Tel.: +49-531-3915217\\
              Fax: +49-531-3918126\\
              \email{j.blum@tu-bs.de} }

\date{Received: date / Accepted: date}
\maketitle

\begin{abstract}
After 25 years of laboratory research on protoplanetary dust agglomeration, a consistent picture of the various processes that involve colliding dust aggregates has emerged. Besides sticking, bouncing and fragmentation, other effects, like, e.g., erosion or mass transfer, have now been extensively studied. Coagulation simulations consistently show that $\rm \mu m$-sized dust grains can grow to mm- to cm-sized aggregates before they encounter the bouncing barrier, whereas sub-$\rm \mu m$-sized water-ice particles can directly grow to planetesimal sizes. For siliceous materials, other processes have to be responsible for turning the dust aggregates into planetesimals. In this article, these processes are discussed, the physical properties of the emerging dusty or icy planetesimals are presented and compared to empirical evidence from within and without the Solar System. In conclusion, the formation of planetesimals by a gravitational collapse of dust ``pebbles'' seems the most likely.
\end{abstract}

\section{\label{sec:intro}Introduction}
It is generally assumed that dust in protoplanetary discs (PPDs) grows from initial sizes of $\sim 0.1-1 ~ \rm \mu m$ to planetesimals. The latter are objects of sizes in the range $\sim 1 - 1000$~km and possess sufficient gravitational attraction to allow further accretion of more material in subsequent collisions so that they finally end up in planetary embryos and planets. For bodies smaller than planetesimals, gravitational accretion is not possible in two-body collisions, because typical collision velocities exceed by far their gravitational escape speed. Thus, planetesimals play an essential intermediate role in planet formation. Mutual collisions among protoplanetary dust particles are important for our understanding of the build-up of larger bodies. 

Based upon the pioneering work of \citet{weidenschilling1977a}, \citet{weidenschilling1993} present a comprehensive overview of relative particle motion in the disc. For individual dust grains or small aggregates with diameters smaller than $\sim 10-100 \rm ~ \mu m$, Brownian motion is the dominating source for collisions. Typical collision velocities in this regime are $\stackrel{<}{\sim} 1~\rm mm~s^{-1}$. Aggregates larger than $\sim 10-100 \rm ~ \mu m$ are subject to systematic drift motions relative to the gas, which causes relative velocities, and, thus, collisions, among particles of different sizes. One cause for the relative drift between dust and gas lies in the sub-Keplerian orbital velocity of the pressure-supported gas, which typically rotates around the central star slower than with Keplerian velocity by $\Delta v \sim c^2/v_K$, with $c$ and $v_K$ being the local sound speed of the gas and the Keplerian velocity, respectively. As the dust aggregates do not feel the pressure support, their equilibrium orbital velocity would be Keplerian in the absence of a gaseous nebula. However, in the presence of the gas, small dust aggregates with gas-dust coupling times smaller than the orbital time scale are forced by the gas to move with sub-Keplerian velocity and, thus, feel a net inward-directed force, which causes a radial drift velocity that increases with increasing aggregate size. Following \citet{weidenschilling1993}, these radial drift velocities reach a maximum value of $\sim 60~\rm m~s^{-1}$ for bodies with $\sim 1~\rm m$ diameter, which limits the lifetimes of these bodies to $\sim 100$ years at 1 au. In the particular nebula model used by \citet{weidenschilling1993}, the gas-grain coupling time of m-sized aggregates equals the orbital time scale at 1 au. Aggregates larger than about 1 m travel essentially with Keplerian orbital speeds so that they feel a steady headwind of the slower-orbiting gas. As the frictional effect of this headwind depends on the surface-to-mass ratio of the dust aggregates, the resulting inward drift velocity decreases with increasing aggregate size in this size regime. Aggregates with different dust-gas coupling times (i.e. different sizes for a given porosity) thus differentially drift in the radial direction, which causes collisions among these particles. For the same reason, also in the azimuthal (i.e. orbital-velocity) direction, dust aggregates possess relative velocities to one another, which also leads to collisions. An example of the resulting collision velocities can be seen in Fig. \ref{fig:1}. 

Another cause for systematic drift of dust particles relative to the gas is the sedimentary motion of the particles towards the disc midplane. Also here, dust aggregates with larger surface-to-mass ratios sediment slower than those with smaller ones so that collisions among aggregates with different sizes result. This effect is greater for higher elevations above the midplane, due to a higher downward-directed force component and lower gas densities. Naturally, dust particles in the nebula midplane are not affected by sedimentation so that a quantification requires knowledge about the position above the midplane. 

Finally, gas turbulence plays also an important role in the evolution of dust aggregates, because it also causes dust particles to collide. Due to the nature of the turbulence, it causes stochastic particle motion, particularly (but not exclusively) for larger dust aggregates, depending also on the strength of the turbulence. Thus, also dust aggregates of identical aerodynamic properties (or sizes) can collide.  

Dust particles condense as refractory materials (oxides, metals, silicates) in the hot inner disc, semi-volatile materials (carbonaceous and organic matter) in the middle disc, and ices ($\rm H_2O$, $\rm CO_2$, CO, $\rm NH_3$, $\rm CH_4$) in the cold outer disc, respectively. Due to the findings by the \emph{Rosetta} mission, particularly that the cometary nucleus possesses a high D/H ratio \citep{altwegg2015} and contains a high abundance of molecular O$_2$ \citep{bieler2015}, it is also possible that the water ice in the outer parts of the solar nebula never reached temperatures high enough to vaporise it so that cometary water ice might be a pre-solar component of the condensed matter in the Solar System. Formation models of protoplanetary discs predicted this \citep{visser2009}. 

Over the past 25 years, a number of experimental investigations have tried to shed light on the outcomes of protoplanetary dust collisions. This paper summarizes the knowledge gained by these laboratory and microgravity experiments. Moreover, the possible growth mechanisms towards planetesimals, resulting from the collision models, will be presented. Finally, predictions about planetesimal properties from the corresponding growth models will be compared with empirical evidence.

\section{\label{sec:lab} Outcomes of laboratory experiments}
The systematic search for a collision model of protoplanetary dust started with the work by \citet{blum1993} who showed that collisions among mm-sized dust aggregates consisting of $\rm \mu m$- or nm-sized silicate grains do not result in sticking, but rather in bouncing (for impact velocities between 0.15 and $\sim 1~\rm m~s^{-1}$) or fragmentation (for impact velocities $\stackrel{>}{\sim} 1~\rm m~s^{-1}$). Since then, a considerable number of papers have been published that expanded the parameter space to dust (aggregate) sizes between $\sim 1 ~\rm \mu m$ and $\sim 10$~cm and the whole range of size ratios between projectile and target, and impact velocities between $\sim 10^{-3}~\rm m~s^{-1}$ and $\sim 100~\rm m~s^{-1}$, for silicate monomer grains of $\sim 1 ~\rm \mu m$ diameter. The review by \citet{blum2008} summarizes the state of the art about 10 years ago. \citet{guettler2010} then used all available information at that time to derive the first comprehensive collision model for silicate aggregates. Since then, numerous new experiments have been performed, expanding, for instance, the parameter space to water ice \citep{gundlach2015} as well as $\rm CO_2$ ice \citep{musiolik2016a} and $\rm CO_2-H_2O$ ice mixtures \citep{musiolik2016b}. In the following, I will summarize our knowledge about the possible collisional outcomes and their dependence on projectile and target size as well as grain material and monomer size.

\subsection{General collisional outcomes}
I will start with the results for aggregates consisting of $\rm \mu m$-sized $\rm SiO_2$ grains, because for this material, the body of empirical evidence is satisfactory.

\paragraph{Outcomes for similar-sized collision partners.}
When the size ratio between the colliding dust aggregates is not too different from unity, the following general collision outcomes have been identified:

\begin{itemize}

\item {\bf Sticking.} 

When the collision energy is small compared to the van-der-Waals binding energy of the colliding dust aggregates, sticking occurs inevitably if the collisions are at least partially inelastic \citep{dominik1997}. For higher impact energies, the degree of inelasticity determines the fate of the colliding dust aggregates. \citet{guettler2010} identified three processes that lead to the complete transfer of both colliding bodies into a more massive dust aggregate, i.e., hit-and-stick behaviour for very small impact velocities, sticking with deformation/compaction of the aggregates, and deep penetration of a somewhat smaller projectile into a larger target aggregate. The supporting laboratory and microgravity experiments were performed by \citet{blum1998,wurm1998,blum2000a,blum2000b,krause2004,langkowski2008,weidling2012,kothe2013,weidling2015,brisset2016,brisset2017,whizin2017}.

When colliding dust particles or aggregates are in the hit-and-stick regime (for which the impact energy is insufficient to cause rolling of the dust grains upon impact, see \citet{dominik1997}), the corresponding aggregates develop a fractal morphology \citep{dominik1997,kempf1999,blum2000a,krause2004,blum2006}, with a fractal dimension in the range $D_{\rm f} \approx 1.1 \ldots 1.9$, depending on the gas pressure if the collisions are caused by Brownian motion \citep{paszun2006}. For higher impact speeds, the growing aggregates become more compact \citep{dominik1997,blum2000b,paszun2009,wada2008}, but possibly remain fractal with fractal dimensions of $D_{\rm f} \approx 2.5$ \citep{wada2008}. 

A general description of the outcomes in hierarchical coagulation is provided by \citet{dominik2016}.

\item {\bf Bouncing.}

When the amount of energy dissipation during the collision is insufficient to allow sticking, but still not large enough to disrupt the colliding bodies (see below), the collisions result in bouncing. Bouncing was found in a variety of experimental investigations \citep{blum1993,heisselmann2007,weidling2012,kothe2013,landeck2016,brisset2016,brisset2017}. Although bouncing might seem to be important only for halting growth or stretching growth time scales, it turned out that bouncing collisions lead to the gradual compaction of dust aggregates. \citet{weidling2009} experimentally found that the equilibrium filling factor (or packing density) of bouncing aggregates is $\sim 0.36$.

\item {\bf Fragmentation.}

Experiments have also shown that collisions at high speeds lead to the fragmentation of the colliding dust aggregates \citep{blum1993,beitz2011,schraepler2012,deckers2013,bukhari2017}. Recently, \citet{bukhari2017} analysed the influence of the aggregate size, size ratio and impact velocity on the outcome of fragmenting collisions. They could show that the mass ratio of the largest fragment to the original aggregate mass decreases with increasing impact velocity and that also the slope of the power law of the fragment size distribution is velocity dependent. On top of that, the onset of fragmentation, i.e. the transition velocity from bouncing to fragmentation, increases for decreasing aggregate mass. 

\item {\bf Abrasion.}

In recent microgravity experiments on low-velocity collisions in a many-particle system, it was discovered that cm-sized dust aggregates suffer a gradual mass loss, although their collision velocities were smaller than the threshold velocity for fragmentation \citep{kothe2016}. Abrasion was not observed for velocities below $\sim 0.1~\rm m~s^{-1}$ and then gradually increased in strength with increasing velocity. The preliminary results by \citet{kothe2016} indicate that the abrasion efficiency is relatively weak so that aggregates need on the order of 1000 collisions before they are completely destroyed.

\end{itemize}

\paragraph{Outcomes when small projectiles hit large targets.}
When a rather small projectile dust aggregate hits a much larger target aggregate, the following additional collision outcomes may occur:

\begin{itemize}
\item {\bf Mass transfer.}

Above the fragmentation limit of the small projectile aggregate, part of its mass is permanently transferred to the target aggregate so that the latter gains mass. In the past years, many experimental investigations have studied mass transfer \citep{wurm2005,teiser2009a,teiser2009b,guettler2010,teiser2011,beitz2011,meisner2013,deckers2014,bukhari2017}. Typical efficiencies of the mass-transfer process are between a few and $\sim 50$ per cent and the deposited mass is compressed to a volume filling factor of typically 0.3-0.4 \citep{teiser2011,meisner2013}.

\item {\bf Cratering.}

Experiments have shown that in the same impact-velocity range as mass transfer, but for larger projectile aggregate, the target agglomerates lose mass by cratering \citep{wurm2005b,paraskov2007}. Similar to the classical high-velocity-impact cratering effect, the impinging projectile excavates more mass than it transfers to the target, so that the target's mass budget is negative. Cratering can be regarded as a transition process to fragmentation of the target. \citet{bukhari2017} have shown that the (relative) mass loss of the target with increasing impact energy of the projectile is a continuous function of the impact energy. The amount of excavated crater mass per impact can be substantial, with up to 35 times the projectile mass, depending on the strength of the target material \citep{wurm2005b,paraskov2007}.

\item {\bf Erosion.}

Again for similar impact velocities, but much smaller projectile aggregates or dust monomers, the effect of erosion has been identified in the laboratory \citep{schraepler2011,schraepler2017} and by numerical simulations \citep{seizinger2013,krijt2015}. The erosion efficiency increases with increasing impact velocity and decreases with increasing impactor mass, so that there is a smooth transition to cratering. \citet{seizinger2013} argue that the erosion efficiency of monomers and small aggregates is higher than that of large aggregate projectiles, because some of the initially eroded mass gets pushed into the target aggregate again and is, thus, re-accreted by the target in impacts with large aggregates. Erosion has also been observed for impacts of $\rm \mu m$-sized ice particles into ice agglomerates \citep{gundlach2015}.

\end{itemize}

The detailed physics of most of the before-mentioned collisional outcomes, i.e., their dependence on collision velocity, size and size ratio of the dust aggregates, their porosity or monomer-particle size have been compiled in the reviews by \citet{blum2008}  and \citet{guettler2010}. \citet{guettler2010} presented the first complete collision model, including all effects known at the time. They distinguished in their model collisions between compact and highly porous dust aggregates and also differentiated between collisions among similar-sized dust aggregates and those between aggregates of very different sizes, because some of the outcomes only occur above or below a certain size ratio between projectile and target aggregate. Short-versions of this collision model have been developed by, e.g., \citet{windmark2012} and \citet{birnstiel2016}.

\begin{figure}
\includegraphics[width=0.75\textwidth]{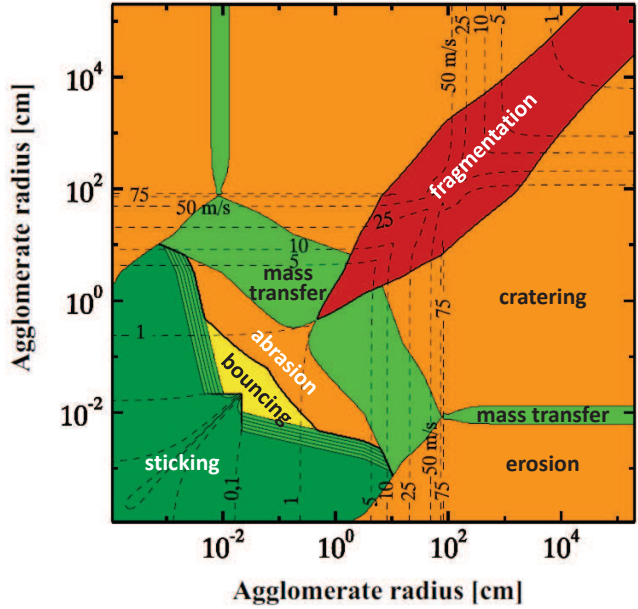}
\caption{Example of the latest collision model under development at TU Braunschweig for $\rm SiO_2$ monomer particles with 0.75~$\rm \mu m$ radius, a minimum mass solar nebula model at 1 AU in the midplane, and a turbulence strength of $\alpha = 10^{-3}$, respectively. Collisional outcomes are marked in colour and labelled. Dashed contours mark the mean collision velocities in units of $\rm m~s^{-1}$. Image credit: Kothe and Blum (in prep.).}
\label{fig:1}      
\end{figure}

The latest experimental findings since the \citet{guettler2010} model will soon be incorporated into a forthcoming dust-aggregate collisions model (see \citet{kothe2016} for a preliminary version). An example of this model is shown in Fig. \ref{fig:1} where on the axes the sizes of the colliding dust aggregates are shown and the coloured regions denote the collisional outcomes. It must be mentioned that the contours and their boundaries can considerably vary with (i) distance to the central star, (ii) PPD model, (iii) turbulence strength, (iv) monomer material, and (v) monomer size (distribution), respectively (see Sect. \ref{sec:mm}).

\subsection{Influence of material and monomer size}\label{sec:mm}
The parameter-space coverage for dust aggregates consisting of micrometre-sized silica monomer grains is quite advanced so that the resulting collision model may be considered relatively reliable. However, the situation is much worse if we consider other abundant materials or nanometre-size constituents of the aggregates. While the latter case can be scaled using dust-collision models, e.g. by \citet{dominik1997,wada2008,lorek2017}, other grain materials require new experimental results at realistic temperatures. Some progress in this field has been made for micrometre-sized water-ice grains and aggregates thereof. Experiments by \citet{gundlach2015} indicate that the sticking-to-bouncing threshold of icy monomer grains is a factor $\sim 10$ higher than the corresponding value for silica grain. The same factor of $\sim 10$ can be found for the bouncing-erosion threshold \citep{gundlach2015}. It must be mentioned that \citet{gundlach2015} also found that the stickiness of small water-ice grains heavily depends on temperature. While for temperatures below $\sim 200$~K the sticking threshold is constant at  $\sim 10 \rm ~ m~s^{-1}$, its value increases steadily for temperatures above $\sim 200$~K. It has recently be shown by \citet{gaertner2017} that the reason for this behaviour is the increase in thickness of a diffuse surface layer around the ice particles for temperatures exceeding $\sim 200$~K. Thus, laboratory experiments on the study of protoplanetary ice aggregation have to be performed at low ambient temperatures and vacuum conditions  \citep{gaertner2017}.

That the material properties may have an enormous impact on the growth behaviour of protoplanetary dust has long been speculated. \citet{kouchi2002,kudo2002} performed low-velocity impact experiments with ``interstellar organic matter analogs''. They found that in the right temperature regime (around $\sim 250$~K), this material possesses an enhanced stickiness. However, for lower and higher temperatures, the material was either too hard or not viscous enough to allow for sufficient sticking. \citet{flynn2013} found that individual monomer grains of a chondritic porous interplanetary dust particle of possible cometary origin were coated with a $\sim 100$~nm thick organic mantle. They estimated the tensile strength of the coated $\rm \mu m$-sized grains to be at least several hundred Pa, comparable to or even higher than the values measured for silica particles. Thus, we may expect that organic coatings may aid planetesimal formation through sticking collisions. However, much more work in this field is required before the role of organic matter has been solved.

Besides water ice, there is only little experimental data on the collision outcome of micrometre-sized particles and their aggregates consisting of other materials. Recently, first laboratory experiments on the collision properties of $\rm CO_2$ aggregates have been reported by \citet{musiolik2016a}. It seems that their sticking threshold does not differ from that of $\rm SiO_2$.

\section{\label{sec:formationofplanetesimals} Pathways to planetesimals}
Based on the above described collision model for protoplanetary dust aggregates, three pathways to planetesimals seem to be feasible, which will be described, including their pros and cons, in the following.

\subsection{\label{sec:GC} Formation of planetesimals by the gentle gravitational collapse of a concentrated cloud of dust ``pebbles''}
If the majority of the dust grains consist of siliceous material, Monte-Carlo simulations of the protoplanetary aggregation process have shown that after an initial process of fractal growth, sticking and bouncing collisions lead to the formation of rather compact mm- to cm-sized aggregates with volume filling factors of $\sim 0.36$ within $\sim 10^4$ orbital time scales \citep{zsom2010}. Further growth is inhibited by the bouncing barrier and no other growth process (i.e., mass transfer) can be reached. It turned out that the maximum aggregate size depends on (i) the PPD model, with a minimum mass solar nebula model resulting in the biggest aggregates \citep{zsom2010}, (ii) the distance to the central star and (iii) the composition of the dust \citep{lorek2017}. Increasing stellar distances and increasing dust-to-ice ratios result is smaller maximum aggregate sizes \citep{lorek2017}. Moreover, the aggregates are fluffier further away from the central star, for smaller monomer grains and if the dust-to-ice ratio is low \citep{lorek2017}. In the literature, the dust aggregates in this stage of growth are termed ``pebbles''. The presence of mm- to dm-sized ``pebbles'' (and partially the absence of larger particles) was confirmed in a number of PPD observations 
\citep{vanboekel2004,dalessio2006,natta2007,birnstiel2010,ricci2010,perez2012,trotta2013,testi2014,perez2015,tazzari2016,liu2017}.

Under the conditions that 

\begin{enumerate}
\item the Stokes numbers $\rm St$ of the ``pebbles'' are in the range $\mathrm{St} \sim 10^{-3} \ldots 5$ \citep{yang2017},

\item the ``metallicity'' of the PPD is larger than a Stokes-number-dependent threshold value, and 

\item the local dust-to-gas mass ratio is above unity,
\end{enumerate}

the streaming instability (SI, see below and \citet{youdin2005} for the original work) can further concentrate the ``pebble'' cloud until a gravitational collapse leads to the formation of planetesimals. Here, the Stokes number is defined by $\mathrm{St}= \tau_f \Omega$, with $\tau_f$ and $\Omega$ being the gas-dust friction time and the orbital frequency, respectively. The ``metallicity'' is defined by the vertically-integrated dust-to-gas ratio. A relation between the minimum required ``metallicity'' and the Stokes number of the aggregates can be found in \citet{yang2017}. An absolute minimum ``metallicity'' of $\sim 0.015$ is required for $\mathrm{St} \sim 0.1$ to allow the SI to work. 

The SI is a collective-particle effect in which the frictional feedback from the solid to the gaseous component of the disc is taken into account. In contrast to a single ``pebble'', a highly-concentrated region of ``pebbles'' with a local dust-to-gas mass ratio above unity has two major effects: (i) Due to the mutual aerodynamic shielding, the effective cross section of the region is much smaller than for all dust ``pebbles'' combined. This means that the SI region experiences less headwind and, thus, radially drifts much slower and azimuthally much faster (see Sect. \ref{sec:intro}). By this, it can catch up with all individual ``pebbles'' on its orbit (and with those that reach its orbit during their radial inward drift), which are then incorporated into the SI region by which its mass grows. (ii) Owing to the large mass concentration, the SI region accelerates the ambient gas, which further reduces friction and yields a local minimum in radial drift. Numerical simulations have shown that this can lead to the gravitational collapse of the ``pebble'' cloud \citep{johansen2007}.

\citet{lorek2017} showed that the known collision properties of dust aggregates indeed lead to the formation of ``pebbles'' with sufficiently large Stokes numbers. Numerical work on the SI has resulted in predictions of the mass-frequency distribution function of planetesimals, which can be fitted by a power law with exponent $-1.6 \pm 0.1$ \citep{simon2016,schaefer2017,simon2017}.

\subsubsection{Benefits}
The formation of planetesimals by the SI avoids any problems that arise for dust aggregates larger than the ``pebble'' size, i.e., the bouncing barrier for aggregate sizes $\stackrel{>}{\sim}$ mm-cm \citep{zsom2010}, the fragmentation barrier for impact velocities $\stackrel{>}{\sim} 1 \rm ~ m~s^{-1}$ \citep{bukhari2017}, the erosion barrier \citep{schraepler2017}, or the drift barrier \citep{weidenschilling1977a}. On top of that, the SI is a fast process. As soon as the conditions (see above) are met, the gravitational collapse occurs on the order of tens to a few thousand orbital time scales \citep{johansen2007,yang2017}.

\subsubsection{Problems}
As already stated above, the SI requires a minimum ``metallicity'' and relatively high spatial pre-concentration of the ``pebbles''. As it is difficult to reach the ``optimal'' Stokes number for the SI ($\mathrm{St} \sim 0.1$), they can reach a ``minimal'' Stokes number of $\mathrm{St} \sim 1.5 \times 10^{-3}$, which requires a ``metallicity'' of 0.03 or higher \citep{lorek2017}. In turn, such a high ``metallicity'' requires a partial dissipation of the gaseous PPD of initial solar ``metallicity'' so that planetesimal formation can only occur at later stages of the disc phase. To reach the required pre-concentration of the ``pebbles'', various ideas have been suggested, including concentration inside or between turbulence eddies or concentration in pressure bumps (see, e.g., \citet{johansen2014}).

\subsection{\label{sec:MT} Formation of planetesimals by collisional growth}
An alternative approach to planetesimal formation has been described by \citet{windmark2012b,windmark2012,garaud2013,booth2018}. After reaching the bouncing barrier, the gap to fragmentation, mass transfer or cratering (see Fig. \ref{fig:1}) is overcome by, e.g., assuming a velocity distribution of the dust aggregates. Once new small projectile aggregates are formed in the destructive processes, mass transfer allows some of the dust aggregates to grow to planetesimal sizes.

\subsubsection{Benefits}
The processes of fragmentation, cratering, and mass transfer have been empirically proven to exist for dust aggregates so that they can be regarded as robust. Assumptions of velocity distributions seem reasonable, because turbulence-induced velocities have a stochastic nature and variations in the mass density of aggregates will lead to variations in drift velocities for aggregates of the same mass. At 1 au, 100-m-sized aggregates may form within $1 \ldots 5 \times 10^4$ years \citep{windmark2012b,garaud2013}.

\subsubsection{Problems}
Although the time scales for growth to the 100-metre level are reasonably short at 1 au, they are much longer further out in the disc, and the maximum aggregate sizes are thus considerably smaller. \citet{garaud2013} showed that at 30 au, even after $6 \times 10^5$ years, the maximum aggregate size is only a few metres. Typically, the growth time scales are so long that radial drift (outside dust traps) limits the maximum size achievable \citep{booth2018}. Recently, \citet{schraepler2017} showed that even under the most favourable conditions of growth everywhere in the parameter space shown in Fig. \ref{fig:1}, except for the part marked ``erosion'', no aggregates larger than $\sim 0.1$ m can form, due to the overwhelming and self-supporting effect of erosion.

\subsection{\label{sec:IG} Formation of planetesimals consisting of sub-micrometre-sized water-ice particles}
The biggest obstacles to direct collisional growth to planetesimals are (i) the low velocity for the transition from sticking to bouncing, which limits the ``pebble'' size before the bouncing barrier is hit, and (ii) the low collision energies required for compaction, which results in fast radial drift time scales. To overcome both, smaller and stickier monomer grains can be used \citep{wada2009,okuzumi2012,kataoka2013}. In their planetesimal-formation model, \citet{kataoka2013} assume that the material is dominated by the relatively sticky water ice \citep{gundlach2015} and that the ice grains possess radii of 0.1~$\rm \mu m$, for which compaction is harder to achieve than for 1~$\rm \mu m$-sized grains. They calculate the collisional growth path of the resulting ice aggregates at 5 au and 8 au. Following an initial fractal growth phase, which leads to minimal mass densities of $\sim 10^{-5}~\rm g~cm^{-3}$, the aggregates then are compacted in mutual collisions, by the ram pressure of the gas and due to self-gravity. At the end of the simulation, the icy aggregates possess radii of $\sim 10$~km and mass densities of $\sim 0.1~\rm g~cm^{-3}$. 

\subsubsection{Benefits}
Due to the choice of very small ice particles with high restructuring impedance and high collision threshold, the bouncing barrier is never reached. The growth time scales to planetesimals are very short ($\sim 10^4$ years), due to the high porosity and, thus, high capture cross section of the fluffy aggregates \citep{krijt2015}. As a consequence, radial drift is negligible. Results have been confirmed by \citet{krijt2015} and \citet{lorek2017}.

\subsubsection{Problems}
Relaxing one of the assumptions (high stickiness, high resistance to compaction) does not lead to planetesimal sizes, and the resulting (smaller) aggregates suffer considerable radial drifts. The role of erosion for the survival of icy agglomerates was studied by \citet{krijt2015}, who found that erosion can limit growth if the erosion threshold is smaller than typical relative velocities between small and large aggregates. Empirical data for sub-micron-sized ice particles are still missing. On top of that, the collisional physics of highly fluffy (ice) aggregates has not been studied so far so that judging the collisional outcomes is difficult.

\section{\label{sec:planetesimalproperties} Properties of planetesimal formed in the different scenarios}
Based on our knowledge about the collision physics of dust and ice aggregates, all of the three pathways to planetesimals (see Sect. \ref{sec:formationofplanetesimals}) are in principle feasible (but mind their individual problems, as stated in Sect. \ref{sec:formationofplanetesimals}). Obviously, we need more laboratory experiments and refined collision models to assess whether or not any of the three formation models is really capable of predicting the formation of planetesimals. 

Alternatively, we may consider the predictions of the diverse models about the physical properties of the resulting planetesimals to better assess the likeliness of the respective planetesimal-formation scenario. Several distinct physical quantities can be estimated with which the three models can be distinguished. These comprise the size of the resulting planetesimals, the volume filling factor (i.e., the fraction of total planetesimal volume actually occupied by matter), the tensile strength (i.e., the internal cohesion of the material), the collisional strength (i.e. the energy required to fragment the colliding bodies such that the biggest surviving mass equals half the original mass), the Knudsen diffusivity (i.e., the resistance to gas flow), and the thermal conductivity, respectively. Table \ref{tab:comprop} compiles estimates of these values for the three formation models and expands on earlier approaches by \citet{blum2006b,blum2014}. Please mind that the values shown in Table \ref{tab:comprop} for the gravitational-collapse model are only valid for small planetesimals, because only then the ``pebbles'' survive the gravitational collapse intact \citep{wahlberg2014,wahlberg2017}. However, the approach used by \citet{wahlberg2014,wahlberg2017} ignores feedback of the collapsing dust particles to the gas cloud. A more rigorous treatment of the collapsing two-phase-flow problem was performed by \citet{shariff2015}, but the implications to the fate of the ``pebbles'' have not been considered yet. In any event, the average lithostatic pressure of a body with $R=50$~km in radius and an average mass density $\rho = 1000 ~\mathrm{kg~m^{-3}}$ is $p=\frac{4}{15} \pi G \rho^2 R^2 = 1.4 \times 10^5$~Pa, with $G$ being the gravitational constant, which exceeds the crushing strength of the pebbles. Thus, for  planetesimals larger than $\sim 10$~km in this formation model, the physical values should approach those of the mass-transfer model.

\begin{table}
\caption{Comparison between the three formation scenarios of planetesimals described in Sect. \ref{sec:formationofplanetesimals}. }
\label{tab:comprop}
\begin{tabular}{lcccc}
\hline\noalign{\smallskip}
        & Gravitational & Mass & Icy \\
        & collapse & transfer & agglomerates & \\
        & (Sect. \ref{sec:GC}) & (Sect. \ref{sec:MT}) & (Sect. \ref{sec:IG}) \\
\noalign{\smallskip}\hline\noalign{\smallskip}
        Size of planetesimals [km] & $\stackrel{<}{\sim} 1000$ [1] & $\stackrel{<}{\sim} 1$ [2-4]& $\sim 10$ [5]\\
        \hline
        Volume filling factor & $0.36 \times 0.6 \approx 0.2$ [6-7] & $\sim 0.4$ [8]& $\sim 0.1$ [5] \\
        &$\sim 0.4$ $^*$&&\\
        \hline
        Tensile strength of \\
        interior [Pa] & $\sim 1-10$ [9-10]& $\sim 10^3-10^4$ [8,11]& $\sim 10^3-10^4$ (guess)\\
        \hline
        Critical fragmentation \\
        energy for 1 m-sized \\
        body $[\rm J~kg^{-1}]$ & $\sim 10^{-5}$ [12]& $\sim 10^{2} [12]$ & $\sim 10^{2}$ [12]\\
        \hline
        Normalised Knudsen \\
        diffusivity & $\equiv 1$ & $\sim 10^{-4} \ldots 10^{-3}$ [13]& $\sim 10^{-5} \ldots 10^{-4}$ [13]\\
        \hline
        Thermal conductivity & $10^{-3}-1$ [14]& $10^{-2}-10^{-1}$ [14]& $10^{-2}-10^{-1}$ [14]\\
        $\rm [W m^{-1} K^{-1}]$ & (conduction/radiation) & (conduction) & (conduction)  \\
\noalign{\smallskip}\hline
\multicolumn{4}{l}{References:}\\
\multicolumn{4}{l}{[1] \citet{schaefer2017}, [2] \citet{windmark2012}, [3] \citet{windmark2012b},} \\
\multicolumn{4}{l}{[4] \citet{garaud2013}, [5] \citet{kataoka2013}, [6] \citet{weidling2009},}\\
\multicolumn{4}{l}{[7] \citet{zsom2010}, [8] \citet{kothe2010}, [9] \citet{skorov2012},}\\
\multicolumn{4}{l}{[10] \citet{blum2014}, [11] \citet{blum2006b}, [12] \citet{krivov2018},}\\
\multicolumn{4}{l}{[13] \citet{gundlach2011b}, [14] \citet{gundlach2012}}\\
{$^*$ For planetesimals with $R \stackrel{>}{\sim} 10-50$~km}
\end{tabular}
\end{table}

A word of caution is necessary before we use Table \ref{tab:comprop} for a comparison between the various formation scenarios. While the gravitational-collapse and the mass-transfer model are physically distinct, this is only partly the case for the mass-transfer and the icy-agglomerates model. Both scenarios rely on the intrinsic ``stickiness'' of the grains, but vary only in the material and size of the dust/ice grains and, thus, potentially in the region of the protoplanetary disc where they can be applied. 

It is evident from Table \ref{tab:comprop} that in principle some of the listed parameters are sensitive enough to be used to distinguish between the various formation models. Exceptions might be the thermal conductivity, for which the uncertainties and overlaps of the different models are too large, and the volume filling factor for large planetesimals, which is affected by lithostatic compression. Uncertainties in the quantification of the thermal conductivity, the tensile strength and the fragmentation energy arise from their strong dependencies on material properties and temperature. The latter becomes of utmost importance for the thermal conductivity of planetesimals formed through a gravitational collapse, because here radiative heat transport, which is intrinsically strongly temperature dependent, may dominate over conduction \citep{blum2017}.

\section{\label{sec:empevi} Empirical evidence}
We have seen above that the three formation models of planetesimals allow predictions, which can be compared to empirical evidence in our own Solar System or in extrasolar planetary system. Below, several criteria are listed, which allow to draw conclusions on the formation process of planetesimals.

\begin{enumerate}
\item Size-frequency distribution in the asteroid and Kuiper belt.

Both, the asteroid belt and the Kuiper belt exhibit a ``knee'' in the size-frequency distribution around $\sim 100$ km. Asteroids smaller than that size are collisional fragments, whereas larger bodies are assumed to be primordial planetesimals. This may indicate that planetesimals were (on average) born big \citep{morbidelli2009}, but  \citet{weidenschilling2011} pointed out that the current size distribution in the asteroid belt can also be reproduced with sub-km-sized planetesimals.

\item Debris discs.

Debris discs are thought to represent the dusty end of a collision cascade among planetesimals. Thus, modelling the collision processes and fitting them to the observed debris-disc brightnesses may provide information about the planetesimal properties in these extrasolar pre-planetary systems. Recently, \citet{krivov2018} modelled the collisional cascades within debris discs for both, the gravitational-collapse and the mass-transfer scenario of planetesimal formation. As a matter of fact, \citet{krivov2018} found agreement between observed and predicted debris-disc brightnesses for both models. However, due to the different collisional strengths (see Table \ref{tab:comprop}), the number of small (sub-km-sized) bodies in the two scenarios is very different. Although these bodies are unobservable in debris discs, our own Solar System might provide a clue to whether the mass-transfer or the gravitational-collapse model more likely represent the true formation scenario. \citet{krivov2018} point out that only the latter is in agreement with the observed low number of sub-km-sized bodies in the Solar System.

\item Fractal particles in comet 67P.

The discovery of fractal particles in comet 67P-Churyumov-Gerasimenko (hereafter 67P) by two \emph{Rosetta} instruments \citep{fulle2015,fulle2016b,mannel2016} can only be explained if the comet consists of larger entities between which the primordial fractal dust aggregates are captured \citep{fulle2017b}. In the mass-transfer model, the relatively high impact velocities would certainly destroy the pebbles and render them into a compact configuration. The same is true for the icy-planetesimal formation model, which does not provide sufficiently large (cm-sized) void spaces to store the fractal particles until today.

\item Physical properties of comet 67P.

It was argued earlier that comets can only be formed through the gravitational-collapse process, because their dust activity requires a gas pressure below the dry dust layer that exceeds the sum of cohesion and gravitational force of the dust \citep{kuehrt1994,kuehrt1996,skorov2012,blum2014,gundlach2015c,gundlach2016}. With the \emph{Rosetta} mission, a much more detailed investigation on the make-up of cometary nuclei became possible. Recently, \citet{blum2017} showed that observations of the cometary properties by various \emph{Rosetta} and \emph{Philae} instruments (sub-surface and surface temperatures, size-frequency distribution of surface and coma dust, tensile strength) point towards comet 67P consisting of ``pebbles'' with 3-6 mm radii. 

\item General properties of comets.

\citet{blum2017} also showed that a formation of comet 67P by the gravitational collapse of a ``pebble'' cloud correctly predicts the porosity and continued dust activity of the comet. 

As already discussed before (see Table 3 in \citet{blum2006b}), comets are highly porous objects. Experiments revealed that bouncing dust ``pebbles'' possess a volume filling factor of $\sim 0.36$ \citep{weidling2009,zsom2010}. Concentrating these dust aggregates by a (gentle) gravitational collapse results in a packing density of $\sim 0.6$ \citep{onoda1990}. Combining these two filling factors results in the overall packing density of $\sim 0.22$, which is compatible with estimates for comet 67P \citep{kofman2015,paetzold2016,fulle2016a}. The two alternative models fail to predict this value (see Table \ref{tab:comprop}). 

To sustain gas and dust activity of comets over many apparitions, a process is required that quasi-continuously emits gas and dust in the mass-proportion present inside the comet nucleus. To achieve this, the sub-surface gas pressure of the sublimating ices has to overcome the tensile strength of the surface material \citep{kuehrt1994,kuehrt1996,skorov2012,blum2014,gundlach2015c,gundlach2016}. Only the hierarchical architecture of the comet following its formation through the gravitational collapse of a cloud of ``pebbles'' predicts sufficiently small tensile strengths (see Table \ref{tab:comprop}) to allow dust activity for realistic temperatures of the sub-surface ices \citep{skorov2012,blum2014,gundlach2015c,gundlach2016}. As soon as the gas pressure exceeds the tensile strength of the overlaying dust layer, the latter is torn off and transported away. Due to the now free access to the ambient vacuum, the increased local outgassing rate decreases the ice temperature so that ice sublimation slows down and the (local) pressure becomes insufficient to continue dust emission, until a sufficiently thick dust layer has formed again, whereupon the process repeats.

\item Evidence for ``pebbles'' in other comets.

The flyby of the EPOXI spacecraft by comet 103P/Hartley revealed ``pebbles'' of the right size to have formed the comet by the streaming instability and a subsequent gravitational collapse \citep{kretke2015}. Moreover, comets leave trails of dust along their orbits with particle sizes large enough to not be subjected to considerable radiation pressure. A recent analysis of the trail profiles of comet 67P indicates dust ``pebbles'' of millimetre sizes to be the dominant constituents of the trail \citep{soja2015}, in agreement with earlier estimates by \citet{agarwal2010,fulle2010}.

\end{enumerate}

\section{Conclusion}

In this paper, I showed in Sect. \ref{sec:lab} that the current state-of-the-art of pre-planetary dust-aggregate collisions is such that the initial dust growth by sticking collisions leads to mm- to cm-sized aggregates (``pebbles'') for siliceous materials and monomer sizes of $\sim 1~\rm \mu m$. At this stage, inter-aggregate bouncing becomes and important effect, which decreases the porosity of the ``pebbles'' until they reach filling factors of $\sim 0.36$.

Two scenarios can explain the formation of planetesimals out of these ``pebbles'': (i) the gravitational-collapse scenario, which requires a high spatial concentration of the ``pebbles'' (e.g., by the streaming instability) and a subsequent gravitational collapse, or (ii) the mass-transfer scenario, in which the aggregates further grow by high-velocity collisions between small projectile aggregates and larger target aggregates; here, the projectiles disintegrate upon impact and transfer part of their mass to the target. Pros and cons of the two models are discussed in Sect. \ref{sec:formationofplanetesimals}.

For sub-micrometre-sized water-ice monomers, a third scenario for the formation of planetesimals is possible, which relies on the high sticking and restructuring thresholds of very small ice grains. Also for this model, pros and cons are discussed in Sect. \ref{sec:formationofplanetesimals}.

The planetesimals resulting from the three formation models possess very different properties, which were shown in Table \ref{tab:comprop} in Sect. \ref{sec:planetesimalproperties}. These properties in principle allow a comparison to ``real'' planetesimals. Such a comparison was made in Sect. \ref{sec:empevi} where I compiled pieces of evidence that suggest that planetesimals consist of macroscopic dust ``pebbles''. The ``knee'' in the asteroid size distribution, the paucity of sub-km-sized bodies in the Solar System, the discovery of primordial fractal particles on comet 67P as well as many general physical properties of comets support the conjecture that planetesimals formed by a smooth gravitational collapse of dust ``pebbles'', following a spatial concentration of dust aggregates by the streaming instability.

Although the \emph{Rosetta} mission to comet 67P has provided us with many new insights into cometary nuclei, we are far away from a comprehensive and closed picture of their formation, although, as stated above, the gravitational collapse of a ``pebble'' cloud, seems to be a viable assumption. First of all, it is unclear whether the gravitational collapse can lead to km-sized objects at all or produces initially much larger bodies. In the latter case, cometary nuclei must be collisional fragments, which seem to have morphologically and compositionally survived the break-up of their parent bodies quite intact. However, even if planetesimals form small (and by whatever process), it is unclear how they could have survived without undergoing catastrophic collisions \citep{morbidelli2015}. Clearly, more work on the formation and (collisional) evolution of planetesimals needs to be done.

\begin{acknowledgements}
I thank the Deutsche Forschungsgemeinschaft and the Deutsches Zentrum f\"ur Luft- und Raumfahrt for continuous support. I thank Stefan Kothe for providing me with Fig. \ref{fig:1}. I also thank Stu Weidenschilling for his constructive suggestions during the review process. Finally, I thank ISSI for inviting me to a wonderful and fruitful workshop.
\end{acknowledgements}

\bibliographystyle{aps-nameyear}      
\bibliography{sample}                

\end{document}